\documentstyle[prc,aps,twocolumn,epsfig]{revtex}
\begin{document}
\draft

\twocolumn[\hsize\textwidth\columnwidth\hsize\csname
@twocolumnfalse\endcsname

\title{\bf How to model Bose-Einstein correlations}
\author{O.V.Utyuzh$^{1}$, G.Wilk$^{1}$ and Z.W\l odarczyk$^{2}$}
\address{$^{1}$The Andrzej So\l tan Institute for Nuclear Studies,
Ho\.za 69; 00-689 Warsaw, Poland}
\address{$^{2}$Institute of Physics, Pedagogical University,
Konopnickiej 15; 25-405 Kielce, Poland}
\date{\today}
\maketitle

\begin{abstract}
We propose new algorithm for numerical modelling of Bose-Einstein
correlations (BEC). It is based on the fact that identical particles
subjected to BEC do, by definition, bunch themselves in a maximal
possible way, restricted only by conservation laws, in the same cells
in phase-space.
\end{abstract}
\vspace{5mm}

]

The study of Bose-Einstein correlations (BEC) \cite{HBT,GGLP} has
already long history \cite{BEC}. It is motivated mainly by their
ability to provide space-time information about multiparticle
production processes. This is of particular importance in searches
for proper dynamical evolution of heavy ion collisions, especially
with respect to the production of Quark Gluon Plasma (QGP)
\cite{QGP}. Because of their complexity the multiparticle production
processes are modelled using Monte Carlo event generators of
different sorts \cite{QGP,GEN}. Their totally probabilistic
structure excludes {\it a priori} genuine BEC, which are of purely
quantum statistical origin. One can only hope to {\it model} BEC by
changing the original output of these generators in such a way as to
reproduce measured experimentally two-particle correlation function,
\begin{equation}
C_2(Q=|p_i - p_j| )\, =\,
             \frac{N_2(p_i,p_j)}{N_1(p_i)\, N_1(p_j)}, \label{eq:C2}
\end{equation}
defined as the ratio of the two-particle distributions to the product
of the single-particle distributions. This is achieved by a suitable
bunching finally produced identical particles (mostly pions of the
same sign) in phase-space and can be performed in many ways. In 
\cite{LS} it is done locally by shifting slightly (pulling closer to
each other) momenta of like-charge mesons according to some
prescribed weights and correcting afterwards for the energy-momentum 
imbalance introduced this way. Other method is to select events
already showing such bunching and count them as many times as
necessary in order to get desirable $C_2(Q)$. This is achieved by
using special weights calculated for each event from the outcome of
the event generator used \cite{AR,FW,Wibig}. Such procedure preserves
energy-momentum balance in each event but changes final
single-particle distributions obtained from the event generator and
should be corrected accordingly. Because in practical terms it is 
prohibitively time consuming, sometimes this method is reduced to the
simple multiplication of particle configuration recorded in each
event by such weight (to this cathegory belongs also "afterburner"
method \cite{AFTER,GEHW}). However, in this case energy-momentum
balance is again altered and should be corrected accordingly.

In all these approaches parameters describing appropriate weights do
not reflect in any way the possible structure of the hadronizing
source. They have been chosen simply to describe the experimental
$C_2$ function (\ref{eq:C2}). In this letter we would like to propose
another method of introducing desired bunching of identical
particles. It will make direct use of the fact that identical
particles subjected to BEC have, by definition, very strong tendency
to bunching themselves in a maximal possible way (restricted only by
conservation laws, especially by the energy-momentum conservation) in
the same cells in phase-space \cite{BSWW}. Usually event generator
provides us in each event with a number of charged and neutral
secondaries (assumed to be pions, for simplicity) of which we know
their energy-momenta and sometimes also spacio-temporal positions of
their production points. The weighting procedure mentioned above is
just a kind of filtering of only those events which, by shear chance,
are already showing desired pattern of BEC. Performing such filtering
already in each event would result in BEC pattern preserving at the
same time both the energy-momenta and total multiplicity
distributions. This is possible if one resignes from some information
provided by event generator. The energy-momenta and total charges
cannot be changed because they are directly measured. However, the
spacio-temporal pattern of the event or the charge distribution among
particles in the event can be altered as not directly observable.
Changes in spacio-temporal pattern corresponds in a sense to
introduction of quantum mechanical element of uncertainty to the
otherwise classical event generator and has been discussed to some
extend in \cite{QUANT,GEHW,AR}. We shall concentrate therefore in
what follows on charge allocation to the produced particles. In
particular, we shall propose to subject all identical particles
produced in a given event to a kind of specific charge assignement
filter, which either allocates charges to particles, if event
generator does not perform this function, or otherwise changes their
original charge allocation. This is quite different approach from
that usually discussed \cite{HBT,GGLP,BEC}, because instead of
symmetrization of appropriate multiparticle wave function one works
in the number of particles basis and implements bosonic character of
the produced secondaries by bunching the like-ones in the phase
space. In \cite{BSWW} it was done by introducing concept of
elementary emmitting cell groupping particles of the same charge
({\it EEC}), in \cite{OMT} by using a new statistical model based on
the information theory approach. In any case (following the original
HBT effect \cite{HBT}) one has to model quantity measuring
correlations of fluctuations present in the system:    
\begin{eqnarray}
\langle n_1 n_2\rangle &=&  \langle n_1\rangle \langle n_2\rangle
                     + \langle \left(n_1 - \langle n_1\rangle\right)
               \left(n_2 - \langle n_2\rangle\right)\rangle \nonumber\\
                     &=& \langle n_1\rangle \langle n_2\rangle
                         + \rho \sigma(n_1)\sigma(n_2). \label{eq:COV}
\end{eqnarray}
Here $\sigma(n)$ are dispersions of the multiplicity distribution $P(n)$
and $\rho$ is the correlation coefficient depending on the type of
the particles produced: $\rho = +1,-1,0$ for bosons, fermions
and Boltzmann statistics, respectively. The proposed algorithm should
then provide us with $C_2(Q)$, which can be written in the form:
\begin{eqnarray}
C_2(Q=|p_i-p_j|) &=& \frac{\langle
n_i\left(p_i\right)n_j\left(p_j\right)\rangle}
  {\langle n_i\left(p_i\right)\rangle\langle
n_j\left(p_j\right)\rangle}\nonumber\\
       &=& 1 + \rho \frac{\sigma\left(n_i\right)}
                             {\langle n_i\left(p_i\right)\rangle}
                        \frac{\sigma\left(n_j\right)}
                        {\langle n_j\left(p_j\right)\rangle} .
\label{eq:algor}
\end{eqnarray}
It means that BEC can be regarded as a measure of correlation of
fluctuations. To get $\rho > 0$ it is enough to select one of the
produced particles, allocate to it some charge, and then allocate (in
some prescribe way) the same charge to as many particles located near
it in the phase space as possible. In this way one forms a cell in 
phase-space occupied by like-charged particles only. This process
should be repeated untill all particles are used. Notice that,
contrary to all previous methods of modelling BEC, neither the
original energy-momentum distributions nor the spacio-temporal
pattern provided by our event generator are altered. On the other 
hand, we change completely the already existing charge pattern 
but retain both the initial charge of the system and its total
multiplicity distribution. Therefore this method works only when we
can resign from controling the charge flow during hadronization
process. The procedure of formation of such cells will be controlled
by appropriate weights deciding whether or not a given neighbour of
the initially selected particle should be counted as its another
member. Using selection procedure which leads to a geometrical
particle distribution in cells, one maximizes second term in
(\ref{eq:algor}) because in this case $\sigma = <n>$. This is the
general idea of what should be done.  

The proposed algoritm is then as follows: let in the $l^{th}$ event
our generator provides us with $n_l = n_l^{(+)} + n_l^{(-)} +
n_l^{(0)}$ particles. Keeping their energy-momenta $\{p_j\}$ and
spacio-temporal positions $\{x_i\}$ intact, we allocate to them
charges in the following way:
\begin{itemize}
\item[$(1)$] The {\it SIGN} is choosen randomly from: "+", "-"or "0"
pool, with weights proportional to $p^{(+)}_l=n_l^{(+)}/n_l$,
$p^{(-)}_l = n_l^{(-)}/n_l$ and $p^{(0)}_l=n_l^{(0)}/n_l$. It is
attached to particle $(i)$ chosen randomly from the particles
produced in this event and not yet reassigned new charges.
\item[$(2)$]  Distances in momenta, $\delta_{ij}(p) = \left|p_i -
p_j\right| $, between the chosen particle $(i)$ and all other
particles $(j)$ still without signs are calculated and arranged in
ascending order with $j=1$ denoting the nearest neighbour of particle
$(i)$. To each $\delta_{ij}(p)$ an appropriate weight $P(i,j)\in
(0,1)$ is then assigned (the form of which will be discussed below). 
\item[$(3)$] A random number $r \in (0,1)$ is selected from a uniform
distribution. If $n^{SIGN}_l > 0$, i.e., if there are still particles
of given {\it SIGN} with not reassigned charges, one checks the
particles $(j)$ in ascending order of $j$ and if $r < P(i,j)$ then
charge {\it SIGN} is assigned also to the particle $(j)$, the
original multiplicity of particles with this {\it SIGN} is reduced 
by one, $n^{SIGN}_l = n^{SIGN}_l -1$, and the next particle is selected:
$(j)\Rightarrow (j+1)$. However, if  $r> P(i,j)$ or $n^{SIGN}_l =0$
then one  returns to point $(1)$ with the updated values of
$p_l^{(+)},~p_l^{(-)}$ and $p_l^{(0)}$. Procedure finishes when
$n_l^{(+)} = n_l^{(-)} = n_l^{(0)} = 0$, in which case  one proceeds
to the next event. 
\end{itemize}

It is easy to check that this algoritm indeed leads to a geometrical
(Bose-Einstein) distribution of particles in the phase-space cells
formed in this procedure \cite{FOOT1} accounting therefore for their
bosonic character (Bose-Einstein statistics). That this leads to
$C_2(Q) > 1$ for $Q\rightarrow 0$ was already demonstrated in
statistical model based on information theory (with conservation of
charges imposed) \cite{OMT}. In contrast to \cite{OMT} case we are
working with particles already provided us by some event generator
and both the sizes of phase-space cells and their number are varying
from event to event and within given event depending on the values of
weights $P(ij)$. It is important to realize that, because we do not
restrict {\it a priori} the number of particles which can be put in
a given cell, we are automatically getting BEC of {\it all orders}.
It means that $C_2(Q=0)$ calculated in such environmet of
multiparticle BEC can exceed $2$ (cf., \cite{Wibig}). 

To demonstrate abilities of our algoritm we show in Fig. 1 results
for two different models of hadronization of mass $M$: CAS and MaxEnt
(limited for simplicity to one-dimensional cases only). CAS is based
on the phase-space and space-time cascade model discussed by us
recently \cite{CAS}, MaxEnt is simple statistical model based on
information theory concepts \cite{EIWW} providing us only with the
momenta distributions. None of them shows BEC. As can be seen, our
algoritm results in a clear BEC type behaviour of correlation
functions $C_2(Q)$ (which come out to be of exponential shape) even
with constant weights $P(i,j)=0.5$. They apparently do not depend
much on the type of hadronization used. The comparison between models
is done in the following way: to each CAS event characterized by the
multiplicity $n_l$ one builds the corresponding MaxEnt event
according to the procedure outlined in \cite{EIWW} (calculating the 
corresponding Lagrange multiplier $\beta_l$ or "temperature"
$T_l=1/\beta_l$, which describes distribution of particles in phase
space). Using now the same multiplicities, $n_l, \, n^{(+)}_l, \,
n^{(-)}_l$ and $n^{(0)}_l$, as in CAS one calculates the
corresponding BEC in MaxEnt. 

\begin{figure}
\noindent
\centerline{\epsfig{file=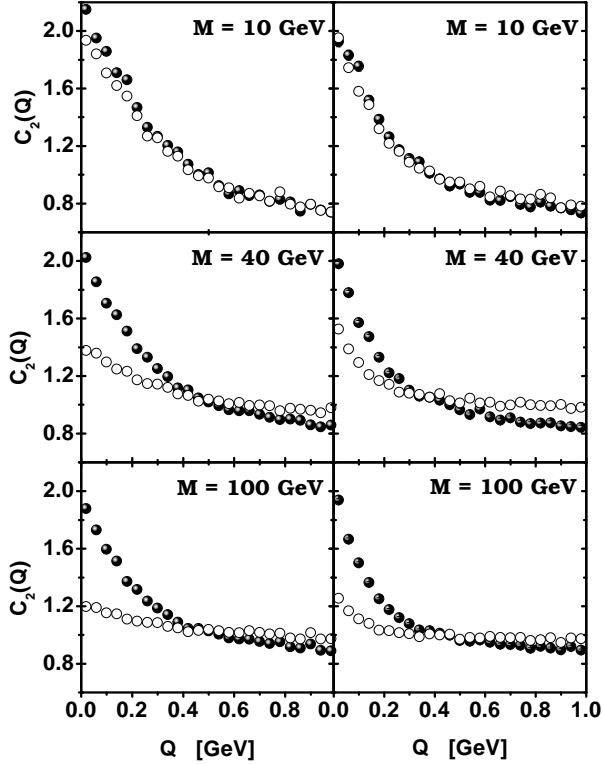, width=80mm}}
\caption{Comparison of BEC obtained using our algoritm for CAS 
(left panels) and MaxEnt (right panels) types of the emmiting 
one-dimensional sources of different masses $M$. Black symbols 
are for $P=0.5$ and open symbols for the "most natural" choices 
of $P$ discussed in text.}
\label{Plot1}
\end{figure}

The only additional place where dynamics can enter are weights
$P(ij)$. Leaving aside more detailed analysis of this problem, we
shall display here, in addition to the constant weigths arbitrary
chosen to be equal $P(ij)=0.5$, also a kind of the "most natural"
choices of these weights, which uses only the available information
provided by event generator:
\begin{equation}
P(ij) = \exp\left[- \frac{1}{2} \delta^2_{ij}(x)\cdot\delta^2_{ij}(p)
\right] \label{eq:CAS}
\end{equation}
for CAS (particles $(ij)$ described by spacio-temporal wave packets
would have widths given by their momentum separation
$\delta_{ij}(p)$) and
\begin{equation}
P(ij) =\exp\left[-\, \frac{\delta^2_{ij}(p)}{2\mu_T T_l} \right]
\label{eq:MaxEnt}
\end{equation}
for MaxEnt ($\mu_T$ is transverse mass put here to be equal $0.3$ GeV
and the role of spatial dimension is now played by the "temperature"
$T_l$ of the $l^{th}$ event mentioned above). The BEC effect is now
weaker, but evidently present. 

The similarity of BEC patterns in both types of models
originates in the fact that BEC effect is in our case given entirely
by the number of particles of the same charge in a phase-space cell.
This depends on $P$, the bigger $P$ the more particles and bigger
$C_2(Q=0)$; smaller $P$ leads to the increasing number of cells,
which, in turn, results in decreasing $C_2(Q=0)$, as already noticed
in \cite{BSWW}. Only after connecting $P$ with details of
hadronization process, some differences between models start to be
visible. They originate entirely in differences in the number of
elementary cells and number of particles located in them. Therefore
the "sizes" $R$ obtained from the exponential fits to results in Fig.
1 (like $C_2(Q) \sim 1 + \lambda\cdot \exp(-Q\cdot R)$ where
$\lambda$ being usually called chaoticity parameter \cite{BEC})
correspond to the sizes of respective elementary cells rather than to
sizes of the hadronizing sources itself. For $P=0.5$ the "size" $R$
varies weakly between $0.66$ to $0.87$ fm from $M=10$ to $100$ GeV
for CAS and between $0.83$ and $1.32$ fm for MaxEnt whereas for other
weights discussed above it varies, respectively, from $0.60$ to
$0.57$ fm for CAS and from $0.94$ to $1.67$ fm MaxEnt. This should be
contrasted with the real sizes of CAS sources changing from $0.29$ fm
for $M=10$ GeV to $1.61$ fm for $M=100$ GeV. 

\begin{figure}
\noindent
\centerline{\epsfig{file=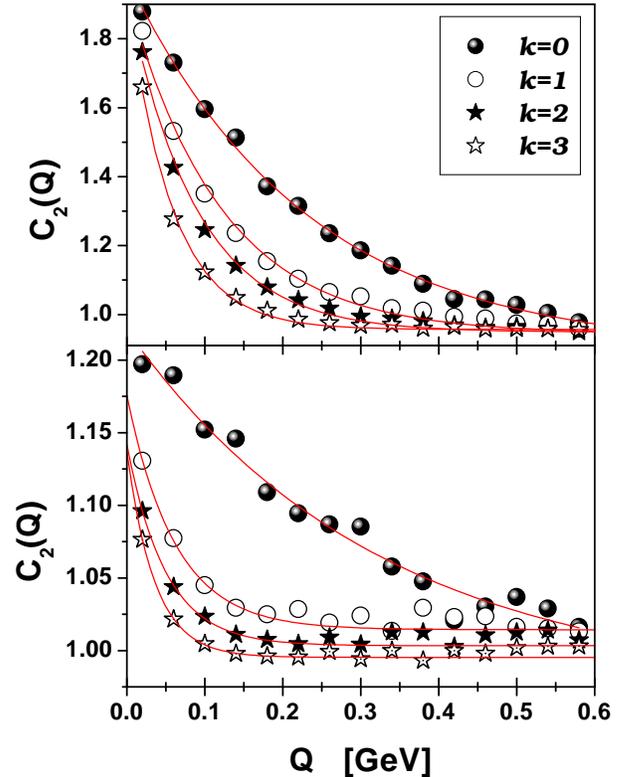, width=80mm}}
\caption{Dependence of BEC calculated using our algorithm for the CAS
type of the hadronic source on the actual number of sources emmiting
particles: the original mass $M=100$ GeV ($k=0$) is divided into
$2^k$ sources with $k=1,2,3$, respectively, positioned in the same
place. Upper panel is for $P=0.5$ and lower for the "most natural"
choice of $P$ discussed in text.} 
\label{Plot2}
\end{figure}

So far we were considering only single sources. Suppose now that mass
$M$ consists of a number of sources hadronizing independently. It
turns out that resulting $C_2$'s are sensitive to whether one applies
our algorithm of assigning charges to all particles from subsources
taken together or to each of the subsource independently. Whereas the
later case results in the similar "sizes" $R$ (defined as before)
with $C_2(Q=0) = 1+\lambda$ falling dramatically with increasing $k$
(roughly like $1/2^k$ subsources), the former leads to rougly the same
$C_2(Q=0)$ but the "size" $R$ is now increasing. The results for this
case are shown in Fig. 2. The corresponding $R_k$ are now equal to,
respectivly, $0.87$ fm, $1.29$ fm, $1.99$ fm and $3.35$ fm for
$P=0.5$ and $0.57$ fm, $3.26$ fm, $4.01$ and $5.59$ fm for gaussian
$P$ as defined above \cite{FOOT2}.  

To summarize: we propose new and simple method of numerical modelling
of BEC. It is based on the expected phase-space behaviour of
identical bosons, conserves the energy-momenta and does not alter
spacio-temporal pattern of events or any single particle inclusive
distribution. We find that value of $C_2(Q=0)$ (defining chaoticity
parameter $\lambda$) depends on the number of elementary  cells in
the way already discussed in \cite{BSWW} and that "radius" $R$
extracted from the exponential fits is practically independent on
the size of the source, provided it is a single one. In the case when
it is composed of a number of elementary sources, $R$ increases with
their number, unless they are treated independently by our algorithm.
It is because one has in this case higher density of particles what
results in smaller average $Q$, and this in turn leads to bigger $R$.
On the contrary, for the independently treated sources density of
particles subjected to our algorithm does not change, hence the
avarage $Q$ and $R$ remain essentially the same. However, because in
this case the influence of pairs of particles from different
subsources increases, the effective $\lambda = C_2(0)-1$ decreases
now substantially (as was already observed in \cite{BSWW}). It means
that with such algoritm we can attempt to fit experimental data. One
should remember, however, that so far our results are limited to
one-dimensional case only and most probably should be first 
generalized to $3$-dimensional case as well. Also the role of 
resonances (for which charge selection should be done independently)
and final state interactions (in particular Coulomb interactions)
should be first clarified. This will be done elsewhere. Finally, we 
would like to mention that our algorithm leads to strong
intermittency showing up after its application, in a fashion similar
to that discussed already in \cite{Wibig} (enhancing the natural
intermittency pattern existing already in CAS \cite{CAS} and
introducing it in MaxEnt case where without BEC it was not present). 

The partial support of Polish Committee for Scientific Research
(grants 2P03B 011 18 and 621/ E-78/ SPUB/ CERN/P-03/DZ4/99) is
acknowledged.



\begin{references}

 \bibitem{HBT} R.\ Hanbury-Brown and R.\ Q.\ Twiss, {\it Philos.Mag.} {\bf
               45}, 663 (1954); {\it Nature} (London) {\bf 178}, 1046
               (1956).
 \bibitem{GGLP} G.Goldhaber {\it et al.}, {\it Phys. Rev.} {\bf 120},
                300 (1960).
 \bibitem{BEC} Cf. the following reviews for the most recent
               (complementary) presentations of the Bose-Einstein
               correlations: R.\ M.\ Weiner, {\it Phys. Rep.} {\bf 327},
               249 (2000); U.\ A.\ Wiedemann and U.\ Heinz, {\it Phys.
               Rep.} {\bf 319}, 145 (1999);
               T.\ Cs\"org\H{o}, in {\it Particle Production Spanning
               MeV and TeV Energies}, eds. W.Kittel et al., NATO Science
               Series C, Vol. 554, Kluwer Acad. Pub. (2000), p. 203
               (see also: hep-ph/0001233).

 \bibitem{QGP} Cf. proceedings of any Quark Matter conference and
               references therein, for example  QM99, eds. L.Riccati
               et al., {\it Nucl. Phys.} {\bf A661} (1999).

 \bibitem{GEN} See, for example, K.Werner, {\it Tools for RHIC:
               Review of Models}, hep-ph/0009153 and references
               therein.

 \bibitem{LS} L.L\"onnblad and T.Sj\"ostrand, {\it Eur. Phys. J.}
              {\bf C2}, 165 (1998).

 \bibitem{AR} B.Andersson and W.Hoffman, {\it Phys. Lett.} {\bf B169},
              364 (1986); B.Andersson and M.Ringn\'er, {\it Nucl.
              Phys.} {\bf B513}, 627 (1998).

 \bibitem{FW} A.Bia\l as and A.Krzywicki, {\it Phys. Lett.} {\bf B354},
              134 (1995); K.Fia\l kowski and R.Wit, {\it Eur. Phys. J.}
              {\bf C2}, 691 (1998); K.Fia\l kowski, R.Wit and J.Wosiek,
              {\it Phys. Rev.} {\bf D58}, 094013 (1998).
 \bibitem{Wibig} T.Wibig, {\it Phys. Rev.} {\bf D53}, 3586 (1996).
 \bibitem{AFTER} J.P.Sullivan et al., {\it Phys. Rev. Lett.} {\bf 70},
                 3000 (1993).
 \bibitem{GEHW} K.Geiger, J.Ellis, U.Heinz and U.A.Wiedemann,
                {\it Phys. Rev.} {\bf D61}, 054002 (2000).
 \bibitem{BSWW} M.Biyajima, N.Suzuki, G.Wilk and Z.W\l o\-dar\-czyk,
                {\it Phys. Lett.} {\bf B386}, 297 (1996).
                The references to works on effects of such bunching
                on some characteristics of multiparticle production
                are also given there.

 \bibitem{QUANT} H.Merlitz and D.Pelte, {\it Z. Phys.} {\bf A351},
                 187 (1995) and {\it Z. Phys.} {\bf A357}, 175 (1997);
                 U.A.Wiedemann et al., {\it Phys. Rev.} {\bf C56},
                 R614 (1997); T.Cs\"org\H{o} and J.Zim\'anyi,
                 {\it Phys. Rev. Lett.} {\bf 80}, 916 (1998) and {\it
                 Heavy Ion Phys.} {\bf 9}, 241 (1999).
 \bibitem{OMT} T.Osada, M.Maruyama and F.Takagi, {\it Phys. Rev.}
               {\bf D59}, 014024 (1999).
               In this event generator one is choosing particles in
               phase-space cells according to a kind of grand
               canonical distribution with two Lagrange multipliers
               playing roles of "temperature" and "chemical
               potential" and keeping the size of phase-space cells
               containing particles of the same charge as fixed
               parameter of the model.
 \bibitem{FOOT1} For $P(ij)=P=$const its mean multiplicity equals
                 $P/(1-P)$, in fact, because it contains also $n=0$,
                 therefore in the algorithm it will be greater by $1$.

 \bibitem{CAS} O.V.Utyuzh, G.Wilk and Z.W\l odarczyk, {\it Phys. Rev.}
               {\bf D61}, 034007 (2000) and {\it Czech J. Phys.}
               {\bf 50/S2}, 132 (2000) (hep-ph/9910355).

 \bibitem{EIWW} G.Wilk and Z.W\l odarczyk, {\it Phys. Rev.} {\bf D43},
                794 (1991).

 \bibitem{FOOT2} We have checked that these results do not depend 
                 substantially on whether subsources are located at the
                 same point or are separated by $1 \div 2$ fm.

\end{references}
\end{document}